**DD-EbA: An algorithm for determining the number of neighbors in cost estimation by analogy using distance distributions**

**Makrina Viola Kosti, Nikolaos Mittas and Lefteris Angelis**

Department of Informatics, Aristotle University of Thessaloniki 54124, Thessaloniki
GREECE, e-mail: {mkosti, nmittas, lef}@csd.auth.gr}

# Abstract

Case Based Reasoning and particularly Estimation by Analogy, has been used in a number of problem-solving areas, such as cost estimation. Conventional methods, despite the lack of a sound criterion for choosing nearest projects, were based on estimation using a fixed and predetermined number of neighbors from the entire set of historical instances. This approach puts boundaries to the estimation ability of such algorithms, for they do not take into consideration that every project under estimation is unique and requires different handling. The notion of distributions of distances together with a distance metric for distributions help us to adapt the proposed method (we call it DD-EbA) each time to a specific case that is to be estimated without loosing in prediction power or computational cost. The results of this paper show that the proposed technique achieves the above idea in a very efficient way.

# 1 Introduction

The last few years we observe a continuous technological orgasm, which has to do with the importance software has taken in our every day life. As human activity becomes more and more complicated, the need of even more exigent systems seems vital. A demanding system however has demanding requirements regarding the development procedure. A project manager is the one who has to confront the issue in many perspectives, either has it to do with risk management matters, defect prediction etc or effort and time requirements. Effort requirements in means of effort estimation is a task known as Software Cost estimation, which is essential in the early stages of the development and may be performed in any stage of the life cycle of the under estimation project.

To cope with this important task, a plethora of methods has been proposed (Jorgensen and Shepperd, 2007). These estimation methods fall in three main categories.

- *Expert judgment*, in which the calculation of effort is based on human judgment. Although the technique is easy to apply and gives direct evaluation, because of the fact that it relies on instinctive processes is difficult to be in full analyzed (Berger, 1985; Silverman, 1985; Hammond, 1996 and Hogarth, 2005).

- *Estimation by analogy (EbA)*, a form of Case Based Reasoning (Mukhopadhyay et al., 1992). Main aspect of the technique is to use historical projects, the effort of which is known, to estimate a new one.

- *Algorithmic cost estimation*, i.e. COCOMO (Boehm, 1981) and Function Points (Albrecht and Gaffney, 1983). These methods require the application of a cost model, via one or more mathematical equations calculated through statistical data analysis. These approaches, if used correctly and calibrated with historical data, seem to be very useful.

As discussed in Shepperd and Schofield (1997) there are certain advantages that appoint EbA the more attractive in respect with the other systems. In fact, users are more willing to accept solutions from analogy based systems since they are congener to human problem thinking and solving, in direct opposition with the awkward chains of rules or neural sets. As it uses historical projects and the similarity of them to the new project, in order to estimate the effort of the latter, a decision must be made considering what number of neighboring historical projects to use for the estimation. This is a limitation to the method and a variety of solutions has been introduced the last years. One of them, without requiring a predefined number of neighbors, calculates the number of the neighbors that gives the better estimate through the process of Leave – One – Out Cross Validation (LOOCV) (Mittas and Angelis, 2010). The method may define a number for neighbors to choose but the whole procedure is quite time consuming and that number is the same for all the under estimation projects.

Purpose of this paper is to present a new algorithm, which introduces a new notion of distance. It uses the similarity between distributions of distances instead of LOOCV to find the number of appropriate neighbors to be used for estimating new projects. The main characteristic of the proposed method is that it is much less time-consuming with respect to LOOCV EbA, and also adaptive to each project that is to be estimated, since it finds an optimal different number in every application. To explicate the above we discuss in Section 2 about LOOCV EbA, analyzing the steps of the algorithm and afterwards, in Section 3, we introduce our algorithm DD-EbA. Section 4 has to do with the methodology we follow to test the accuracy of our method. In Section 5 we analyze the results of the proposed method and that of LOOCV EbA. Finally, Section 6 consists of a constructive discussion about the results and the new paths this approach opens. Finally, reflections are made about future work.

In this paper, experimentations are based on four well known datasets, namely the Maxwell dataset (Maxwell, 2002), the Desharnais dataset (Desharnais, 1989) and the COCOMO - NASA dataset (PROMISE). All statistical and simulation procedures in this study have been implemented with Matlab programming language.

## 2  Combined Estimation by analogy using LOOCV for determining the number of neighbors

Through the three formerly mentioned effort prediction categories, **E**stimation **b**y **A**nalogy (EbA) has been proposed as a valid technique, starting with Boehm (Boehm, 1981) and has been continuously enriched (Shepperd et al., 1996; Shepperd and Schofield, 1997) and applied on a number of cost data sets.

Main goal of the method is to predict the effort (cost) of the project under estimation (target case) using historical information from completed projects with known effort. After having characterized the new project with attributes (or features) [1] common to the ones of the historical dataset (training set), three basic steps must be followed.

1. Calculation of distances of the target case with the projects of the training set using a distance metric (for example the Euclidean distance),

2. Choice of the nearest neighbors.

3. Effort estimation calculating a statistic (mean, median etc.) and using the efforts of the projects selected in step 2.

An additional advantage of EbA, besides the one mentioned in the introduction, is that it is based on distance measures for finding the nearest neighbors and these measures can be easily calculated even for categorical variables. The limitations of the method have mainly to do with the characterization of the project to be estimated and the choice of the number of neighbors to use (Step 2) for the calculation of the estimate. As pointed out in the description of the method, when a new project arrives it must be inserted in an historical dataset, in which cases are described by a specific number of features. This means that the new project has to be described by exactly the same features. This is not a problem for software organizations that undertake projects in the same domain.

---

[1] The features might be continuous, discrete or categorical and might include the number of interfaces, the level of code reuse, the programming language, the number of LOC etc.

As for the limitation of having to predefine a specific number of neighbors, we already mentioned that LOOCV procedure has widely been applied. In each step of the procedure, a completed project $P_i$ is removed from the dataset and the remaining projects are used for the estimation of its cost YE. The validation of the combined model is based on the YA (actual effort) and the YE (estimated from a predictor) values. The procedure is applied for a range of neighbors $k = [1, \dots, k_{max}]$, that are given as an input for the estimation process. This is adopted to find an optimal number of nearest neighbors to the target case, in order to make the estimation. In the end of every test of $k$, an accuracy measure is calculated (MdAE, see Section 3) so as to chose that specific number of neighbors from the above range, which gives the minimum accuracy. This method is time consuming and lacks of adaption to each specific under estimation project.

## 3   DD-EbA algorithm description

DD-EbA in the present paper focuses on the limitation of the traditional method to choose a different number of neighbors for each target case. The traditional method has no criterion for doing that, which leads to the a priori choice of a fixed number of neighbors, without considering an optimal number for each target case. This means that when a new case has to be estimated, the number of neighbors for the estimation is predefined and fixed.

In this study we try to make a first step towards a new approach, which essentially determines the appropriate number of neighbors, separately for each specific new case to be estimated. This is a way of dynamic calculation of neighbors adapted to the special characteristics of each project under estimation. To achieve that, we assume that in the training set there is a unique project $P_i$ that is similar to the target case $P_{new}$ in the sense that their distances from all the other projects are distributed in a similar way. That is we do not consider the usual similarity measures, but rather the way each project is placed with respect to all the others. The finding of $P_i$ is based on a similarity metric of distance distributions. Once we have found that specific $P_i$, we use the optimal number of its neighbors, i.e. the number that gives the best effort estimate of $P_i$ and we use this number for the estimation of $P_{new}$ by its own neighbors.

Having presented the central idea of the study, we have reached at that point of the research which requests detailed presentation of the algorithm. We rubricate the algorithm in the form of steps which are explicated below.

1. Out of the training set matrix $\mathbf{T}_{nxk}$, we produce a new matrix $\mathbf{D}_{nxn}$ which contains the pairwised distances of the historical cases. Matrix $\mathbf{T}_{nxk}$ comprises $n$ historical projects $(P_1,...,P_n)$, each of them having $k$ attributes $(x_1,...,x_k)$. On the other hand $\mathbf{D}_{nxn}$ comprises the distances amongst the historical projects, with elements $d(i,j)$ being the distance between $P_i$ and $P_j$.

As historical datasets contain various types of variables that have to be treated with a different manner, we have to use a distance metric that takes into account the mixed-type variables. An important procedure is the standardization (between 0 and 1) of each dimension so that every attribute has the same degree of influence and the method is immune to the choice of units. Hence, we used a special dissimilarity coefficient suggested by Kaufman and Rousseeuw (1990). The distance or dissimilarity $d(i,j)$ of projects $P_i$ and $P_j$ which will be computed by their vectors of attributes $\mathbf{X}_i = (x_{i1},...,x_{ik})$ and $\mathbf{X}_j = (x_{j1},...,x_{jk})$, respectively and is given by the following expressions:

$$d(i,j) = \frac{\sum_{m=1}^{p} \delta_{i,j}^{(m)} d_{i,j}^{(m)}}{\sum_{m=1}^{p} \delta_{i,j}^{(m)}} \qquad (1)$$

where:

$$\delta_{i,j}^{(m)} = \begin{cases} 1 \text{ if } x_{im}, x_{jm} \text{ nonmissing} \\ 0 \text{ otherwise} \end{cases} \qquad (2)$$

- If the m-th variable is binary or nominal, the following formula can be used to compute dissimilarities:

$$d_{i,j}^{(m)} = \begin{cases} 1 \text{ if } x_{im} \neq x_{jm} \text{ nonmissing} \\ 0 \text{ if } x_{im} = x_{jm} \end{cases} \qquad (3)$$

- If the m-th variable is of interval or ratio scale, the following formula can be used to compute dissimilarities:

$$d_{i,j}^{(m)} = \frac{\left| x_{im} - x_{jm} \right|}{R_m} \tag{4}$$

where $R_m = \max\limits_{1 \le i \le n}(x_{im}) - \min\limits_{1 \le i \le n}(x_{im})$ is the range of the variable.

- If the m-th variable is ordinal, the dissimilarities are computed as follows:

    i. The values $X_{im}$ are replaced by their ranks $r_{im} \in \{1,...,M_m\}$

    ii. The ranks are transformed to values in the interval $[0,1]$:

    $$z_{im} = \frac{r_{im} - 1}{M_m - 1}$$

    iii. The $z_{im}$ values are treated as interval-scaled.

From the evaluation of the dissimilarity coefficient through Eq. (1), it is clear that there is a special treatment when a dataset has missing values. In these cases, the dissimilarity coefficient does not take into account the feature that contains a missing value (Eq. (2)), while at the same time the vector with the missing values is not completely ignored, i.e. the non-missing information participates in the calculation of the coefficient. Furthermore, the division with the variable range in Eq. (4) ensures that the variables participating in the calculation of the coefficient are standardized.

2. Each row of the distance matrix **D** is considered an empirical distribution, e.g. $f_i$ = [ $d(i,2)$, $d(i,3)$,…, $d(i,\text{n})$] is the i[th] row of matrix **D**, which represents the distances of $P_i$ with all the other projects of the training set and the empirical distribution of distances of that case. Assuming $P_{new}$ is a target case, the effort of which we want to estimate, we calculate the empirical distribution $f_{new}$ = [$d(new,1)$, … , $d(new,n)$] of the distances between the target case and the historical cases. The scope of the above computations is to find the specific historical project that has similar "behavior" with the newly arrived. This essentially means that we want to find that project that has similar placement with respect to all the other projects (and therefore similar distance distribution), with the new one. To find the minor declination of $f_{new}$ from each $f_i$ we use the distance metric used in the two-sample Kolmogorov – Smirnov test (KS test). As it is sensitive to differences in both location and shape of the empirical distribution functions of the two samples, the two-sample KS test is one of the most useful methods for comparing two distributions.

Once we have found the most similar distribution of matrix **D** to the new sample $f_{new}$ we reconsider the training set, making $P_i$, the distance distribution of which was similar to

$f_{new}$, the new target case and the remaining historical cases a new training set. We apply the traditional EbA to this new combination trying a number of neighbors from a predefined range $k = [1,\ldots,k_{max}]$.

3. For each number of neighbors $k$ of Step 2 we calculate the Median Absolute Error (MdAE) and choose that number of neighbors that gives the smaller MdAE. MdAE is a global predictive accuracy measure. We explain accuracy measures at Chapter 4, where we analyze the experimentation. This way we find the number of neighbors by which project $P_i$ is best estimated.

4. Concluding, we use the same number of neighbor to estimate the effort of $P_{new.}$ That is, if the neighbor distribution $f_i$ showed that the optimal number of neighbors is $k$, then the optimal number of neighbors to estimate the effort of $P_{new}$ is also $k$.

In the following chapter we apply DD-EbA to the datasets mentioned in the Introduction of the paper.

# 4   Methodology

Until this point of the paper we analyzed two approaches, LOOCV-EbA and DD-EbA, which can provide cost estimation for new projects. We explained LOOCV-EbA in a nutshell in Section 2 and our approach, the main aspect of which is the dynamic choice of neighbors for EbA, in Section 3.

Our aim now is to examine the predictive power of the two algorithms, by adopting the leave - one - out procedure. Three measures of local accuracy (in the sense that the error is measured in one point) are calculated (Table 1) for each project $i$, $i = 1,\ldots,n$:

1. The magnitude of relative error (MRE)

2. The magnitude of relative error to the estimate (MER)

3. The absolute error (AE)

| $MRE_i = \dfrac{\left|Y_{A_i} - Y_{E_i}\right|}{Y_{A_i}}$ | $MER_i = \dfrac{\left|Y_{A_i} - Y_{E_i}\right|}{Y_{E_i}}$ | $AE_i = \left|Y_{A_i} - Y_{E_i}\right|$ |
|---|---|---|

*Table 1. Local accuracy measures*

The abovementioned local measures are the starting point for the estimation of the global predictive accuracy measures MMRE (mean MRE), MdMRE (median MRE), MMER (mean MER), MdMER (median MER), MAE (mean AE) and MdAE (median AE) (Table 2). Here the term ''global'' is used to show that all local errors are combined to produce an overall measure. The predictive power of each method is presented in Section 5, according to the global accuracy measures that are calculated in the algorithm.

$$MMRE_i = \frac{1}{n}\sum_{i=1}^{n} MRE_i \qquad MMER_i = \frac{1}{n}\sum_{i=1}^{n} MER_i \qquad MdMRE_i = median\{MRE_i\}$$

$$MdMER_i = median\{MER_i\} \qquad MAE_i = \frac{1}{n}\sum_{i=1}^{n} AE_i \qquad MdAE_i = median\{MAE_i\}$$

*Table 2. Global accuracy measures*

Each project is removed from the data set (leave - one - out) and its dependent variable (effort) is estimated by LOOCV-EbA and DD-EbA. After applying this procedure for all the projects, a set of predicted values is generated. We observe then and analyze the results.

Since the cost of each project is predicted by two methods, it is reasonable to use statistical tests for two related samples in order to compare the two cost prediction models on the same data set. The statistical significance for their global predictive accuracy differences can be tested through the non-parametric Wilcoxon signed rank test (Table 3), which tests whether there is a significant difference between the AE of LOOCV-EbA and DD-EbA.

## 5   Experimentation

In order to exploit the predictive power of the methods, the techniques, as mentioned in the Introduction, are evaluated on four real datasets, Maxwell, Desharnais and Cocomo – Nasa. For each data set we examined the accuracy measures derived from nearest neighbors (analogies) that gave the best (lower) MdAE results. In Table 3 we present the results given from the two methods, in means of global accuracy measures.

As we can see from the results there is no significant difference (sig>>0.05) between AE from LOOCV- EbA and DD-EbA. The methods generally perform, in terms of global accuracy measures, the same way. DD-EbA however performs much better in terms of computational cost.

|  |  | MMRE | MdMRE | MMER | MdMER | MAE | MdAE | Sig. |
|---|---|---|---|---|---|---|---|---|
| **Maxwell** | **LOOCV-EbA** | 1.3429 | 0.4983 | 0.6519 | 0.45539 | 5042.4 | 2537.4 | 0.99 |
|  | **DD-EbA** | 1.2059 | 0.5315 | 0.6932 | 0.6198 | 4765.3 | 3363.3 |  |
| **Desharnais** | **LOOCV-EbA** | 0.5708 | 0.4535 | 0.4685 | 0.3837 | 2398.5 | 1356.8 | 0.60 |
|  | **DD-EbA** | 0.6480 | 0.3708 | 0.5075 | 0.3693 | 2342.9 | 1556.8 |  |
| **Cocomo-Nasa** | **LOOCV-EbA** | 0.6758 | 0.3688 | 1.0851 | 0.4044 | 264.31 | 53.4 | 0.72 |
|  | **DD-EbA** | 0.5493 | 0.3504 | 0.8741 | 0.375 | 241.14 | 49.479 |  |

*Table 3. Global accuracy measures – performance of the two techniques*

# 6  Conclusion

In this paper we examine the problem of choosing a specific number of neighbors to estimate cost, when using EbA. In detail, we first analyzed a known from the bibliography method, the LOOCV-EbA. This is a combined technique because apart from traditional EbA, it uses the leave – one out cross validation procedure for each possible number of neighbors and results in a single optimal number of nearest neighbors to be used for new cost estimations. The LOOCV-EbA method works efficiently, but it has a large computational cost since it performs a large number of iterations, growing with the size of the dataset. Furthermore, once the number of neighbors is defined from a training set, all new projects have to be estimated by the same fixed number. Due to this fact, of the inability of LOOCV-EbA to adapt neighbor results to each specific case, we came out with the idea of developing a new algorithm that not only would adapt to each single under estimation case, but also would perform better in terms of computational cost.

DD-EbA is an algorithm that introduces a new notion, the distribution of distances. Scope of the idea was to find the specific historical project that had similar placement, with respect to all the others, with the newly arrived. Once we found that specific historical case, we find the number of neighbors that provide the most accurate prediction of its effort. The new algorithm has the advantage that it does not have to use iteratively the entire training dataset. It operates only on the rows of the derived distance matrix, by comparing them once with the distribution of distances of the single target case. Then, it only uses a sorting of the training effort values to find the optimal number of neighbors. The proposed algorithm avoids the iterative procedure of LOOCV-EbA, saving computational time. We compared the two methods and realized that there was no statistically significant difference between their predictive accuracy. This means that their estimation results were quite alike. DD-EbA though adapts better to the target cases, since it finds a specific number of neighbors for each under estimation case.

The notion of distribution of distances that we used to achieve these results is quite interesting; actually it is the first time it is used in Software Cost Estimation. The first experiments showed us that there is plenty of work to do utilizing it as a tool to find patterns in our results, which could lead us to even better estimations. Finally, we understand that this work has a long way forward and much thought to finally optimize DD-EbA. We see this paper as a preliminary step towards that goal.